\documentclass[twocolumn,twoside,slac_two]{revtex4}
\usepackage{graphicx}
\usepackage{fancyhdr}
\begin{document}

\title{Methods for Measuring the Cosmic-Ray Proton Spectrum With the Fermi LAT}

%

\author{P. D. Smith}
\affiliation{Center for Cosmology and Astro-Particle Physics, Department of Physics, The Ohio State University, Columbus, OH 43210, USA}
\author{R. E. Hughes}
\affiliation{Center for Cosmology and Astro-Particle Physics, Department of Physics, The Ohio State University, Columbus, OH 43210, USA}

\author{B. L. Winer}
\affiliation{Center for Cosmology and Astro-Particle Physics, Department of Physics, The Ohio State University, Columbus, OH 43210, USA}

\author{T. W. Wood, \textit{on behalf of the Fermi LAT Collaboration}}
\affiliation{Center for Cosmology and Astro-Particle Physics, Department of Physics, The Ohio State University, Columbus, OH 43210, USA}

\begin{abstract}
The Fermi Gamma-Ray Space Telescope was launched in June 2008 and the onboard Large Area Telescope (LAT) has been collecting data since August of that same year. The LAT is currently being used to study a wide range of science topics in high-energy astrophysics, one of which is the study of high-energy cosmic rays. The LAT has recently demonstrated its ability to measure cosmic-ray electrons, and the Fermi LAT Collaboration has published a measurement of the high-energy cosmic-ray electron spectrum in the 20 GeV to 1 TeV energy range.  Some methods for performing a similar analysis to measure the cosmic-ray proton spectrum using the LAT will be presented with emphasis on unfolding the reconstructed proton energy.
\end{abstract}

\maketitle

\thispagestyle{fancy}


\section{Introduction}
The Fermi Gamma-ray Space Telescope was designed for the study of many interesting topics in astrophysics ranging from pulsars, active galactic nuclei, gamma-ray bursts, and the indirect detection of dark matter.  Another intriguing subject that has attracted strong interest is the use of the Fermi LAT in the study of cosmic rays.  Early in the design stages of the LAT, the potential for making a cosmic-ray electron measurement was acknowledged \cite{Ormes, Moiseev}. Recently the Fermi LAT Collaboration demonstrated this ability and published a high-statistics measurement of the cosmic-ray electron spectrum in the energy range from 20 GeV to 1 TeV \cite{CREpaper}, containing about 4.5 million events collected over a six month period from August 2008 to January 2009.  The electron spectrum measurement can be used in the study of cosmic-ray propagation and in the constraint of the diffuse gamma-ray emission \cite{CRPropPaper}.  Furthermore, it may also be possible to use the LAT to obtain a measurement of the cosmic-ray proton spectrum, and studies are being conducted within the LAT collaboration to explore this possibility.  We present some preliminary findings illustrating the prospects for such an analysis.

\subsection{The Fermi Gamma-ray Space Telescope}
The Fermi Gamma-ray Space Telescope was launched on June 11, 2008 from Kennedy Space Center.  Two instruments are onboard the satellite.  The main instrument is the Large Area Telescope (LAT), which is a pair conversion telescope sensitive to the energy range from 20 MeV up to greater than 300 GeV.  The second instrument onboard the satellite is the Gamma-ray Burst Monitor (GBM) and is designed to detect GRBs in the 8 keV to 40 MeV energy range.  It consists of 12 NaI and 2 BGO detectors positioned around the spacecraft, and has a large field-of-view allowing it to constantly monitor the entire unocculted sky.  The analyses described here pertain to the LAT instrument, and a more detailed description of the LAT detector can be found in \cite{LATpaper}.

\section{Measuring the Cosmic-ray Proton Spectrum}
\begin{figure}[!t]
\includegraphics[width=80mm]{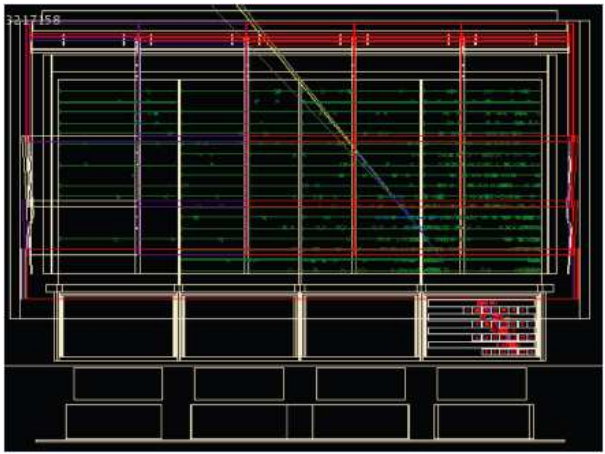}
\caption{A hadron candidate event: a large energy deposit per ACD tile, a small number of extra clusters around the main track; a large number of extra clusters away from the track, and a large and asymmetric shower profile in the calorimeter.}
\label{HadronDisplay}
\end{figure}

\begin{figure}[!t]
\includegraphics[width=80mm]{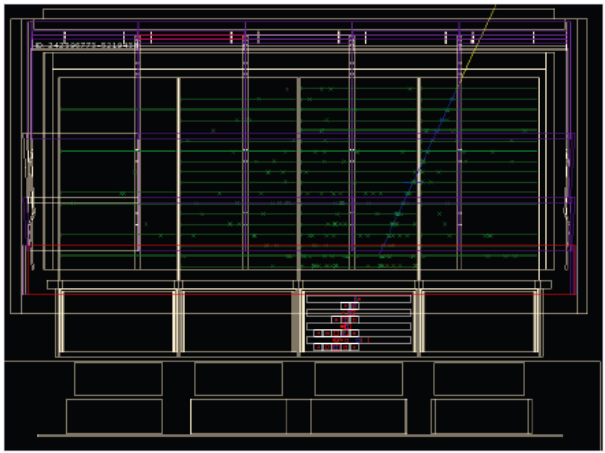}
\caption{An electron candidate event: few ACD tiles hit in conjunction with the track, a clean main track with extra clusters very close to the track (note backsplash from the calorimeter), and a well-defined symmetric shower in the calorimeter (though not fully contained).}
\label{ElectronDisplay}
\end{figure}

We are currently exploring methods to produce a cosmic-ray proton spectrum measurement using the LAT instrument.  However this analysis is inherently more challenging than the electron spectrum measurement.  The main cause for this is that the thickness of the calorimeter is optimized for electromagnetic showers rather than hadronic showers.  Whereas the calorimeter is 8.6 $X_{0}$ on-axis for EM showers, it is only about 0.43 interaction lengths for hadronic showers.  As a result, most protons are only minimum ionizing particles (MIPs) in the LAT, and thus are not useful for the spectrum measurement.  For the protons that do interact and produce a shower, the energy resolution is much worse than for either gamma rays or electrons.  In fact we have found that the reconstructed energy is approximately a lower limit on the true incoming energy.  In addition, the event reconstruction is more complicated than for electrons or gamma rays (Figure \ref{HadronDisplay} and Figure \ref{ElectronDisplay}).  

Here we present our characterization of the instrument response using a preliminary proton event selection applied to events collected via the onboard diagnostic filter (here referred to as the DGN filter).  This filter allows a prescaled sample of all hardware trigger types to be downlinked.  An alternate and independent pathway that is also being utilized for event collection is the high-energy pass feature of the onboard GAMMA filter (to be referred to here as the High Pass filter).  This feature of the onboard software filter allows all events that deposit greater than 20 GeV in the calorimeter to be downlinked.  The High Pass allows for higher statistics than the prescaled sample from the DGN, however with the DGN, there is no lower energy threshold, and the energy spectrum can be extended to lower energies.  These two means of data collection are thus complementary and will also allow a cross-check on the measured energy spectrum.  In the following sections, we further discuss methods for measuring the cosmic-ray proton spectrum using the DGN pathway.  For a earlier discussion using the High Pass feature to measure the proton energy spectrum, refer to \cite{PDSmith}.

\subsection{Preliminary Geometry Factor}
\begin{figure}[!t]
\includegraphics[width=80mm]{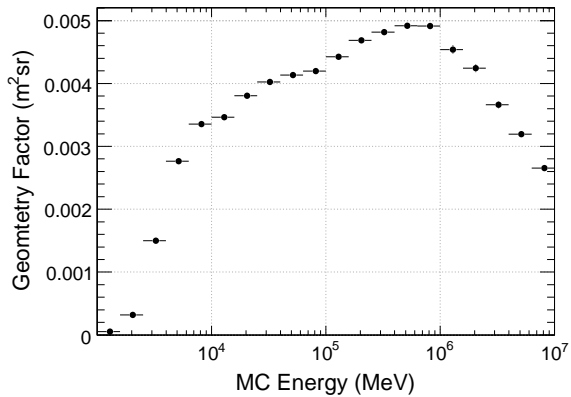}
\caption{The preliminary geometry factor for DGN filter collected events is plotted versus MC energy.}
\label{geomfac}
\end{figure}

A preliminary proton event selection has been developed, and the geometry factor has been calculated from Monte Carlo simulations, with the full instrument geometry, of protons interacting within the LAT.  A preliminary geometry factor as a function of Monte Carlo (MC) energy, or the true incoming particle energy, is shown in Figure \ref{geomfac}.  It peaks at approximately 0.005 m$^2$sr near 700 GeV, and falls off appreciably below 10 GeV.  For comparison, the estimate for events collected via the High Pass using these selections peaks near 0.8 m$^2$sr.  The geometry factor for electrons \cite{CREpaper}, which also uses the High Pass filter, peaks around 2.8 m$^2$sr and further illustrates the fact that most protons are minimum ionizing in the LAT.  However it is worth noting that the preliminary proton geometry factor, using the High Pass, is still comparable to other cosmic-ray experiments, such as the average value of 0.15 m$^2$sr reported by AMS in 2002 \cite{AMS}.

\begin{figure}[!t]
\includegraphics[width=80mm]{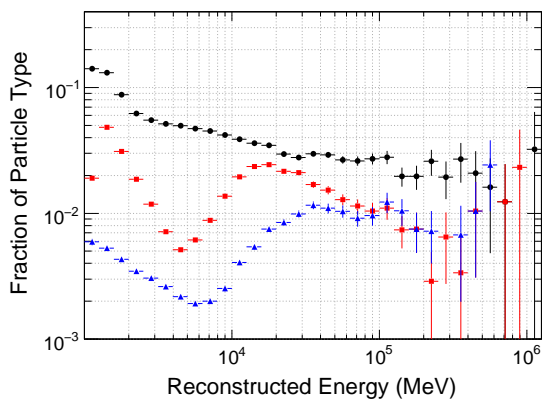}
\caption{The fraction of background contamination for the different background components is plotted versus reconstructed energy using the preliminary proton selections applied to simulated DGN filter collected events.  The fraction of alpha particles is plotted in black, electrons/positrons are plotted in red, and particles with Z$>$2 in blue.  Each background component contributes approximately a few percent or less over most of the energy range.}
\label{bkgcontam}
\end{figure}

\subsection{Preliminary Background Contamination}
For events collected via the DGN filter, we have estimated the background contamination given the preliminary selections using Fermi Monte Carlo simulations of the cosmic-ray environment the LAT encounters, which has been modelled from previous cosmic-ray measurements and again includes the full geometry of the detector.  The dominant background components are found to be electrons/positrons, alphas, and particles with Z$>$2.  The fraction that each particle type contributes to the selected event sample is plotted versus the reconstructed energy in Figure \ref{bkgcontam}.  Over most of the energy range, each background species contributes only a few percent or less to the selected event sample.

\subsection{Energy Unfolding}
\begin{figure}[!t]
\includegraphics[width=80mm]{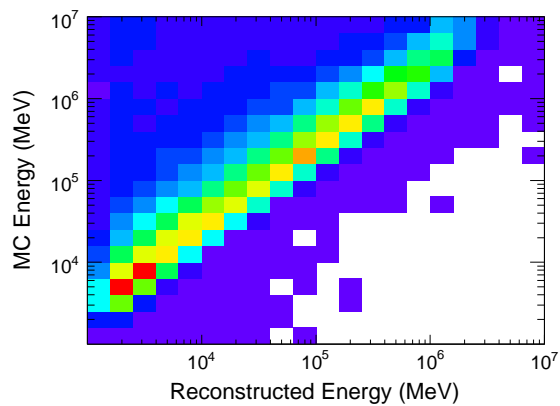}
\caption{The preliminary energy response, for events collected via the DGN filter, is plotted as MC energy versus reconstructed energy.  For a given bin in MC energy, there is a wide range of reconstructed values obtained.  In addition, the reconstructed value can be seen to be approximately a lower limit on the MC energy.  Both illustrate the challenge in measuring the energy of proton events.} \label{eneresp}
\end{figure}

We have also studied the application of an energy unfolding algorithm to the reconstructed energy distribution and the energy response of the instrument given the preliminary selections.  The process of energy unfolding is to calculate a distribution of the true incoming energies of an event sample, given the distribution of reconstructed energies and a detector response matrix.  It should be noted that this procedure does not attempt to correct energies on an event-by-event basis, but rather its goal is to obtain an estimated distribution for the incoming energies of the selected event population.  A similar energy unfolding process was also performed for the LAT electron spectrum measurement, however given that the energy resolution is much worse for protons, the effect of this process on the reconstructed proton spectrum will be much greater.  The energy response used in the unfolding algorithm is plotted in Figure \ref{eneresp}, plotted as MC energy versus reconstructed energy.  This response has been calculated by applying the preliminary proton selections to Fermi LAT Monte Carlo proton simulations using a hard spectrum of E$^{-1}$.  This enables more events to be simulated in the higher-energy bins than would a more typical cosmic-ray (CR) spectrum.  From this response histogram, it can be seen that for a given MC energy bin, there is found to be a wide range of reconstructed energy values.  In addition, it can also be seen that the reconstructed energy is approximately a lower limit on the MC energy.

\begin{figure}[!b]
\includegraphics[width=80mm]{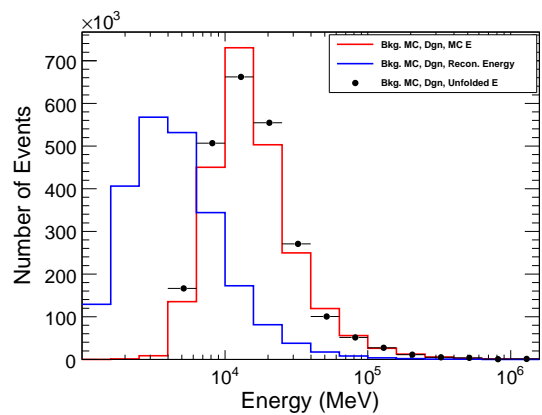}
\caption{Comparison of the unfolded energy distribution (black points) to the MC energy distribution (red line) using preliminary proton selections applied to simulated DGN filter events from Fermi LAT MC cosmic-ray simulations.  The input reconstructed energy distribution (blue line) is also shown.}
\label{eneunfcomp}
\end{figure}

To perform the unfolding procedure, we have used RooUnfold, a ROOT-based framework for unfolding \cite{RooUnfold}.  The unfolding algorithm takes as input the reconstructed energy distribution and the energy response, and the calculated unfolded distribution is returned.  The unfolding procedure is applied to a reconstructed energy distribution taken from a MC sample independent from the one used to create the energy response.  As a test of this procedure, the unfolded energy distribution can be compared to the MC energy distribution.  For this test we have used the reconstructed energy distribution, after preliminary selections, from Fermi LAT MC simulations of the cosmic-ray environment.  It should be noted here that the unfolding procedure is applied here to only the proton component of the event sample surviving the preliminary selections.  The resulting unfolded energy distribution is shown in Figure \ref{eneunfcomp} compared with the MC energy distribution.  The unfolded distribution obtained is a reasonable reproduction of the MC energy distribution with differences of approximately 20\% or less.

\begin{figure}[!t]
\includegraphics[width=80mm]{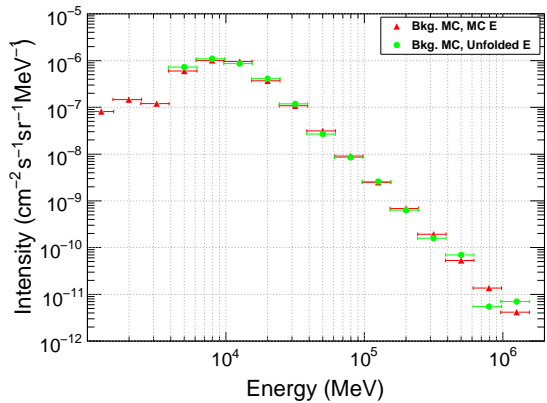}
\caption{The unfolded MC cosmic-ray proton spectrum (green circles) for events from the DGN filter, calculated by dividing the unfolded energy distribution by the simulated livetime and the geometry factor.  For comparison, the proton energy spectrum reconstructed using the MC energy distribution is also shown (red triangles).}
\label{reconMCspec}
\end{figure}

\subsection{Reconstructed Monte Carlo CR Proton Spectrum Using the DGN}
The unfolded distribution shown in Figure \ref{eneunfcomp} can be used to calculate the reconstructed MC cosmic-ray proton spectrum.  The spectrum is calculated by dividing the unfolded distribution by the livetime of the MC simulation used and the geometry factor previously shown.  The resulting spectrum is plotted in Figure \ref{reconMCspec} (green circles).  The reconstructed proton energy spectrum, calculated using the MC energy distribution, is plotted for comparison (red triangles).  Over most of the energy range, the reconstructed MC energy spectrum is reasonably well reproduced by the unfolded energy spectrum.

\subsection{Future Steps in Analysis}
Improved selections are being studied with the goals of spanning a larger energy range, achieving a higher, more constant efficiency, and a more constant background contamination level as a function of energy.  There is also a continuing effort aimed at improving the application of the unfolding procedure to produce a better agreement between the unfolded energy distribution and the MC energy distribution.  The systematic uncertainties associated with the energy unfolding technique are also being estimated.  Furthermore, the background rates need to be estimated and subtracted from the proton candidate event rate, and in addition, the systematic errors from MC/Data discrepancies in the selection variables.  These methods can then be applied to Fermi LAT data and the resulting distribution used to reconstruct a measured cosmic-ray proton spectrum.

\bigskip 

\end{document}